\documentclass{PoS}
\usepackage{amsmath,amssymb,caption,graphicx,subcaption}
\usepackage[utf8]{inputenc}

\newcommand{\pbp}{\langle\bar{\psi}_l\psi_l\rangle}

\title{The curvature of the chiral phase transition line for small values of $\mu_B$}

\ShortTitle{Chiral phase transition curvature}

\author{\speaker{Prasad Hegde} and Heng-Tong Ding (for the Bielefeld-BNL-CCNU collaboration)\\
        Key Laboratory of Quark \& Lepton Physics (MOE) and Institute of Particle Physics,  \\
        Central China Normal University, Wuhan 430079, China.\\
        E-mail: \email{phegde@mail.ccnu.edu.cn \& hengtong.ding@mail.ccnu.edu.cn}}

\abstract{We present preliminary results from an ongoing calculation to determine the curvature of the chiral phase transition line in the chiral limit along the light-light, light-strange and strange-strange chemical potential directions. We do this by studying the appropriate $\mu$-derivatives of the chiral condensate as a function of the quark mass and comparing them to the scaling predictions of $3d$-$O(N)$ theory. We work at a fixed lattice spacing, $N_\tau=6$ and at four different quark masses $m_\pi\approx$ 140, 110, 90 and 80 MeV. For the light quark curvature, we obtain a value 0.03$\leqslant\kappa_{ll}\leqslant$0.11. We also find that both strange and light-strange curvatures are around an order of magnitude smaller. Currently, the light-strange curvature is the least constrained curvature and could have either sign, though our results seem to prefer a slightly negative value.}

\FullConference{The 33rd International Symposium on Lattice Field Theory\\
		14-18 July 2015\\
		Kobe International Conference Center, Kobe, Japan*}

\begin{document}
\section{Introduction}
\label{sec:intro}
The phase diagram of QCD at non-zero quark chemical potential $\mu$ is an open question with relevance to both theory as well as heavy-ion experiments. For physical quark masses and small values of $\mu$, it is known that the chiral transition is a crossover. It is believed that this crossover turns into a second-order phase transition at a point $(T_\text{cp},\mu_\text{cp})$ known as the \emph{critical point}. Indeed, one of the goals of the ongoing Beam Energy Scan (BES) program at RHIC is to find evidence for the existence of the critical point~\cite{Mohanty:2011nm}.

The phase diagram is better understood as a function of quark masses at $\mu=0$. Here, the well-known conclusion of Pisarski and Wilczek ~\cite{Pisarski:1983ms} is that there is a second-order phase transition belonging to the three-dimensional $O(4)$ universality class in the massless two-flavor case. Since then, there has been evidence from lattice QCD~\cite{Ding:2013lfa,Ejiri:2009ac} to support this conclusion, which was originally obtained from calculations with the linear sigma model. 

Since the quark chemical potential does not break chiral symmetry, this second-order transition persists even at $\mu\gtrsim0$. As $\mu$ is increased, this second-order line will eventually meet a first-order transition line at a \emph{tricritical point} $(T_\text{tcp},\mu_\text{tcp})$. If one now introduces a small quark mass, the second-order line shall disappear but the first-order line and tricritical point will remain with the tricritical point turning into the critical point $(T_\text{cp},\mu_\text{cp})$~\cite{Hatta:2002sj}.

Since lattice calculations cannot yet be performed at $\mu \neq 0$, the properties of the phase diagram in this region must be deduced indirectly by extrapolating lattice calculations done at $\mu=0$. One such property is the shape of the second-order line in the chiral limit. For small values of $\mu$, this line can be written as
\begin{equation}
T_c(\mu) = T_c(0)\left[1 - \kappa\left(\frac{\mu}{T_c(\mu)}\right)^2\right] + \mathcal{O}(\mu^4).
\label{eq:definition}
\end{equation}

In this work, we will try to determine the \emph{curvature coefficient} $\kappa$ by taking advantage of the universal nature of second-order phase transitions. Before proceeding, we must mention that there were quite a few results for the curvature coefficient presented at this conference~\cite{Bellwied:2015rza,Cea:2015bxa,Bonati:2015bha}. The coefficient defined there differs from ours in that it was defined for physical quark masses by studying the shift, either of the inflection point in the chiral condensate or of the peak of the chiral susceptibility, as a function of $\mu$~\cite{Endrodi:2011gv,Cea:2015cya,Bonati:2014rfa}. There is some tension between these results and our value for $\kappa$\footnote{However, our current value agrees within error to a previous determination by our group using the same approach but with p4 fermions on $N_\tau=4$ lattices~\cite{Kaczmarek:2011zz}.}. Since the QCD phase transition is a crossover at non-zero quark mass, the value of such a curvature could in principle depend on the observable under consideration. On the other hand, $\kappa$ as defined above is unique since it follows directly from considerations based on universality.

\section{Method}
\label{sec:method}
\begin{table}[!tbh]
\begin{subtable}{0.45\textwidth}
\centering
\begin{tabular}{|c|c|c|c|}            \hline
$m_l/m_s$ & $N_\sigma^3\times N_\tau$ & $\beta$ & \# conf. \\ \hline
1/27 & $24^3\times6$ & 6.025 & 950 \\ \hline 
1/27 & $24^3\times6$ & 6.038 & 1022\\ \hline 
1/27 & $24^3\times6$ & 6.050 & 1650\\ \hline 
1/27 & $24^3\times6$ & 6.062 & 982 \\ \hline 
1/27 & $24^3\times6$ & 6.075 & 1750\\ \hline
\end{tabular}%
\end{subtable}%
\hspace{0.1\textwidth}%
\begin{subtable}{0.45\textwidth}
\centering
\begin{tabular}{|c|c|c|c|}            \hline
$m_l/m_s$ & $N_\sigma^3\times N_\tau$ & $\beta$ & \# conf. \\ \hline
1/40 & $32^3\times6$ & 6.025 &  600\\ \hline 
1/40 & $32^3\times6$ & 6.038 &  600\\ \hline 
1/40 & $32^3\times6$ & 6.050 &  600\\ \hline 
1/40 & $32^3\times6$ & 6.062 &  600\\ \hline 
1/40 & $32^3\times6$ & 6.075 &  600\\ \hline
\end{tabular}%
\end{subtable}
\caption{\small Details of the measurements for the quark masses $m_l=m_s/27$ and $m_l=m_s/40$.\label{tab:1}}
\end{table}

\begin{table}[!bth]
\begin{subtable}{0.45\textwidth}
\centering
\begin{tabular}{|c|c|c|c|}            \hline
$m_l/m_s$ & $N_\sigma^3\times N_\tau$ & $\beta$ & \# conf. \\ \hline
1/60 & $40^3\times6$ & 6.025 & 1210\\ \hline 
1/60 & $40^3\times6$ & 6.038 &  683\\ \hline 
1/60 & $40^3\times6$ & 6.050 & 1208\\ \hline 
1/60 & $40^3\times6$ & 6.075 & 1255\\ \hline
\end{tabular}%
\end{subtable}%
\hspace{0.1\textwidth}%
\begin{subtable}{0.45\textwidth}
\centering
\begin{tabular}{|c|c|c|c|}            \hline
$m_l/m_s$ & $N_\sigma^3\times N_\tau$ & $\beta$ & \# conf. \\ \hline
1/80 & $32^3\times6$ & 6.025 &  992\\ \hline 
1/80 & $32^3\times6$ & 6.038 &  992\\ \hline 
1/80 & $32^3\times6$ & 6.050 & 1000\\ \hline 
1/80 & $32^3\times6$ & 6.062 & 1253\\ \hline 
1/80 & $32^3\times6$ & 6.075 & 1000\\ \hline
\end{tabular}%
\end{subtable}
\caption{\small Details of the measurements for the quark masses $m_l=m_s/60$ and $m_l=m_s/80$.\label{tab:2}}
\end{table}

The properties of the chiral transition can be extracted by comparing QCD observables, obtained using lattice QCD, with the appropriate observables of the $3d$-$O(4)$ model. The fundamental observable in the latter is the magnetization $M$, the net spin per unit volume along the direction of an external magnetic field $H$. This model has a second-order phase transition at $H=0$ and temperature $T=T_c$. Near this phase transition, the magnetization has a non-analytic behavior that is given by
\begin{align}
&& && && M = h^{1/\delta}f_G(z) + f_\text{reg}(T,H) &&\text{with} && z = t/h^{1/\beta\delta}. && && &&
\label{eq:f-mag}
\end{align}
The piece $f_\text{reg}(T,H)$ captures the contribution from the analytic terms. While these are sub-dominant close to the phase transition, their contribution can be significant away from it. The non-analytic behavior is captured by the first term. Specifically, the non-analyticities appear in the prefactor $h^{1/\delta}$ and in the definition of the scaling variable $z$ while the function $f_G(z)$ is an analytic function of $z$. $t$ and $h$ are the reduced temperature and field strength, given by
\begin{align}
&& && && && t = \frac{1}{t_0}\frac{T-T_c}{T_c}, && h = \frac{H}{h_0}. && && &&
\label{eq:scvar}
\end{align}
The QCD analog of $M$ is $m_s\,\pbp/T^4$, where $\pbp$ is the $2$-flavor chiral condensate and $m_s$ is the bare strange quark mass\footnote{The quantity $m_s$ times $\pbp$ is used as it is renormalization-group invariant. The multiplication gets rid of a multiplicative renormalization constant.}. The light-to-strange quark mass ratio $m_l/m_s$ plays the role of the symmetry-breaking field $H$. The change of variables $(T,H)\to(t,h)$ above is suggested by universality theory and is significant only for $f_G(z)$. By universality, this part is the same for QCD as well. We will continue to use the original variables $(T,H)$ for the regular part. More specifically, we will parametrize this part by
\begin{equation}
f_\text{reg}(T,H) = \frac{m_l}{m_s}\left(a_0 + a_1\frac{T-T_c}{T_c}\right).
\label{eq:f-reg}
\end{equation}
We Taylor-expanded $f_\text{reg}(T,H)$ around $(T_c,0)$ and kept the first two terms. This should be sufficient provided that we are not too far from the phase transition. There is an overall multiplicative factor of $m_l/m_s$ since the contribution of the regular part to the chiral condensate vanishes in the chiral limit.


\begin{table}[!t]
\centering
\begin{tabular}{|c|c|c|c|c|}
\hline
$T_c$ [MeV] & $t_0(\times10^{-3})$ & $h_0(\times10^{-6})$ & $a_0$ & $a_1$ \\
\hline
145.6(1) & 3.34(4) & 4.56(4) & 10.1(3) & -384(10) \\
\hline
\end{tabular}
\caption{\small Fit results for the non-universal parameters (From the talk by H.-T.~Ding at this conference~\cite{Ding:2015lat}).}
\label{tab:t0}
\end{table}

\begin{figure}[!ht]
\includegraphics[width=0.45\textwidth]{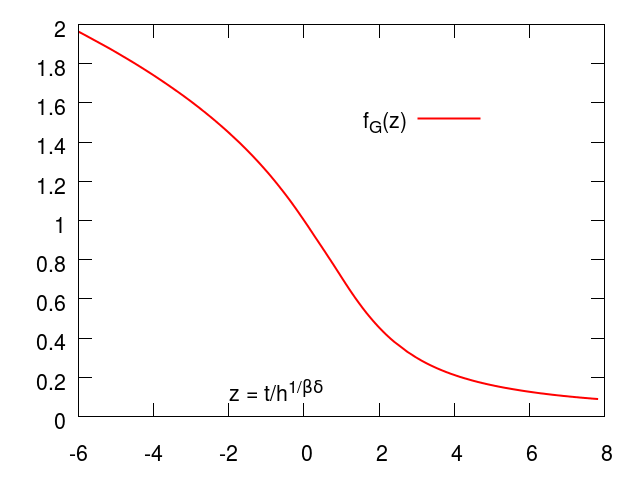}%
\hspace{0.10\textwidth}%
\includegraphics[width=0.45\textwidth]{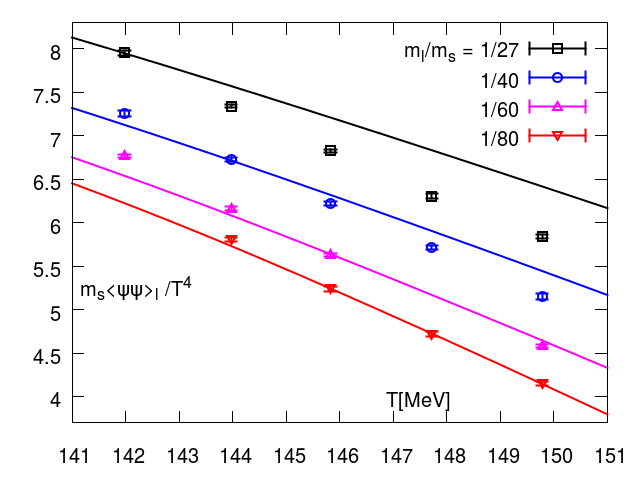}
\caption{\small (Left) The scaling function $f_G(z)$ for the $3d$-$O(2)$ model. (Right) Lattice results for the light chiral condensate.The solid lines are the curves from Eq.~\protect\eqref{eq:f-mag} for each value of $m_l/m_s$. As the quark mass is decreased, the scaling violations get smaller.\label{fig:pbp}}
\end{figure}

The comparison between the $3d$-$O(4)$ model and QCD proceeds by determining the non-universal parameters $t_0$, $h_0$ and $T_c$. Once this is done, the function $f_G(z)$ and its derivatives can be combined to yield predictions from universality theory for various chiral observables. The determination of $t_0$, etc. by our collaboration has been presented elsewhere at this conference~\cite{Ding:2015lat}. These constants were extracted from a joint fit of the scaling relations using $3d$-$O(2)$ critical exponents\footnote{The chiral symmetry group is $O(2)$ rather than $O(4)$ for staggered fermions, if one takes the chiral limit before the continuum limit. In practice, the critical exponents of the two groups are quite similar and the relevant scaling functions $f_G(z)$ too more or less coincide within the $z$ range considered here.} to the chiral condensate and total chiral susceptibility for different quark masses, using $t_0$, $h_0$ and $T_c$ as well as $a_0$ and $a_1$ as the fit parameters. Our results are summarized in Table~\ref{tab:t0}. Fig.~\ref{fig:pbp} shows the agreement between data and fits for the chiral condensate.

\section{Mixed Susceptibility and the Curvature of the Transition Line}
\label{sec:curvature}
The previous discussion did not take into account the quark chemical potential. Unlike the quark mass, a quark chemical potential term $\mu\bar{\psi}\gamma_4\psi$ does not break chiral symmetry. Therefore we expect the scaling relations to remain unchanged. On the other hand, the transition temperature $T_c$ cannot remain the same as before. At $\mu\neq0$, the reduced temperature $t$ becomes $\mu$-dependent:
\begin{align} && && &&
t = \frac{1}{t_0}\left[\frac{T-T_c}{T_c} + \mu^T\mathbf{K}\mu\right], && 
T_c(\mu) = T_c(0)\left[1 - \mu^T\mathbf{K}\mu + \dots\right] && && &&
\end{align}
In the above, $\mu = (\mu_l,\mu_s)^T$, where we have assigned a common chemical potential to the light quarks and another one for the strange quarks. The curvature coefficient $\kappa$ then becomes a $2\times2$ matrix $\mathbf{K}$. Since the transition is always located at $t=0$, we see that the transition temperature $T_c$ must shift as $\mu$ changes. This shift is controlled by $\mathbf{K}$ for small values of the chemical potentials.

\begin{figure}[!bt]
\includegraphics[width=0.45\textwidth]{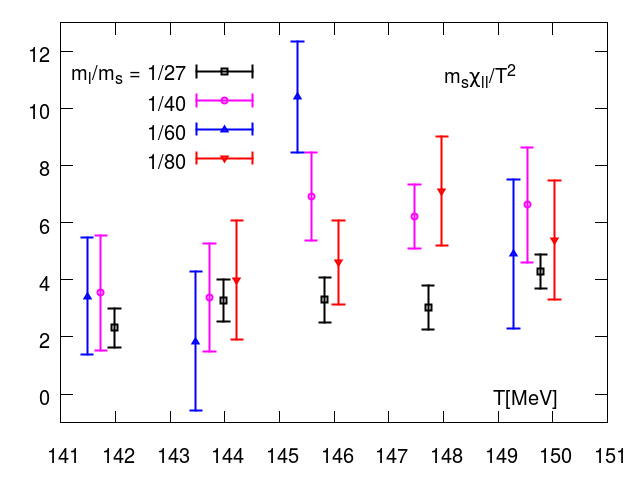}%
\hspace{0.10\textwidth}%
\includegraphics[width=0.45\textwidth]{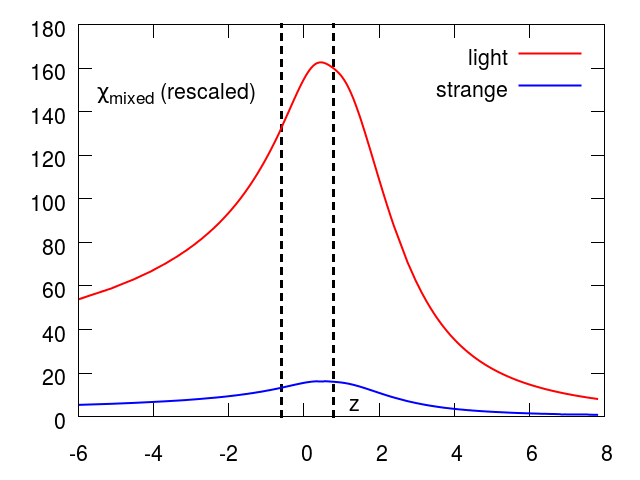}
\caption{\small (Left) Our results for the light mixed susceptibilities(multiplied by $m_s/T^2$) for different quark masses. The $m_l/m_s=1/40$, 1/60 and 1/80 points have been displaced by $-0.25$, $-0.50$ and $+0.25$ MeV along the $T$-axis for clarity. (Right) The scaling function $f'_G(z)$. The different matrix elements appear as multiplicative prefactors to this function. The figure is a schematic depiction of how the peak height is much smaller, and the curve much flatter, when $\kappa$ is small, say for $\kappa_{ss}$ as compared to $\kappa_{ll}$. The black dashed lines show the $z$ range spanned by our lattice measurements.\label{fig:matrix}}
\end{figure}

To determine $\mathbf{K}$, one needs the $\mu$-derivatives of the chiral condensate $\pbp$, known as the \emph{mixed susceptibilities}. According to the scaling formula Eq.~\eqref{eq:f-mag}\footnote{The contribution of the analytic terms to this quantity is small since it is suppressed by $(m_l/m_s)^{0.39}$ ($(m_l/m_s)^{0.34}$ for the $O(4)$ model). We will therefore neglect it in the following.}:
\begin{align}&& && &&
\frac{\partial^2M}{\partial\hat{\mu}_i\partial\hat{\mu}_j} = \frac{2\kappa_{ij}}{t_0}h^{-(1-\beta)/\beta\delta}f_G'(z), && \hat{\mu} = \mu/T. && && &&
\label{eq:mixed}
\end{align}
The function $f_G'(z)$ is plotted in Fig.~\ref{fig:matrix} (right). By calculating the mixed susceptibilities on the lattice and comparing our results to the right-hand side of the above equation, we can determine the only unknowns viz. the curvature coefficients $\kappa_{ij}$'s. 

To calculate the $\mu$-derivatives, we generated $\sim$1000 configurations for four different beta values in the transition region for four quark masses. We have summarized the details of our ensembles in Tables~\ref{tab:1} and~\ref{tab:2}. On each configuration, we calculated the traces of eight operators involving various products of the fermion matrix inverse $D^{-1}$ and matrix derivatives $d^nD/d\mu^n$, $n=1,2$. These were calculated stochastically using Gaussian random noise vectors. By chaining these operators, we were able to compute all eight traces using only four inversions per quark flavor and noise vector. Initially, we calculated these traces using 400 random sources for each operator. Beyond that, we sought to improve the signal by going after the noisiest operator, $\text{tr}(D^{-1}\,dD/d\mu)$, which we calculated using a separate set of 1,000 random vectors. In addition to precise trace estimates however, calculating $\mu$-derivatives of the fermion matrix also requires significant statistics, especially at lighter quark masses. Currently, our estimates are limited by statistics. Work to improve these numbers is currently ongoing.

\begin{figure}[!ht]
\includegraphics[width=0.45\textwidth]{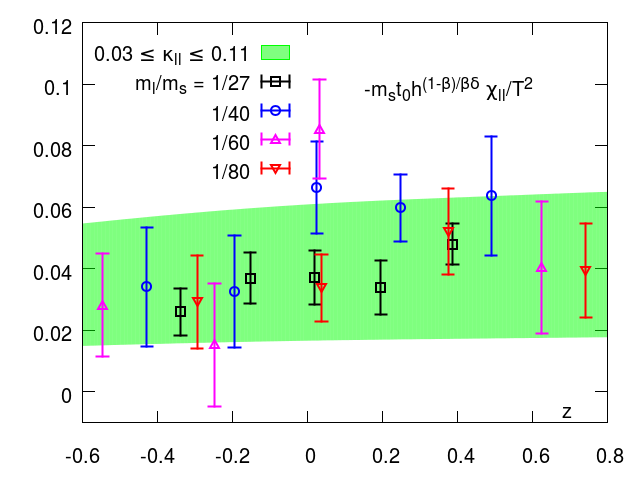}%
\hspace{0.10\textwidth}%
\includegraphics[width=0.45\textwidth]{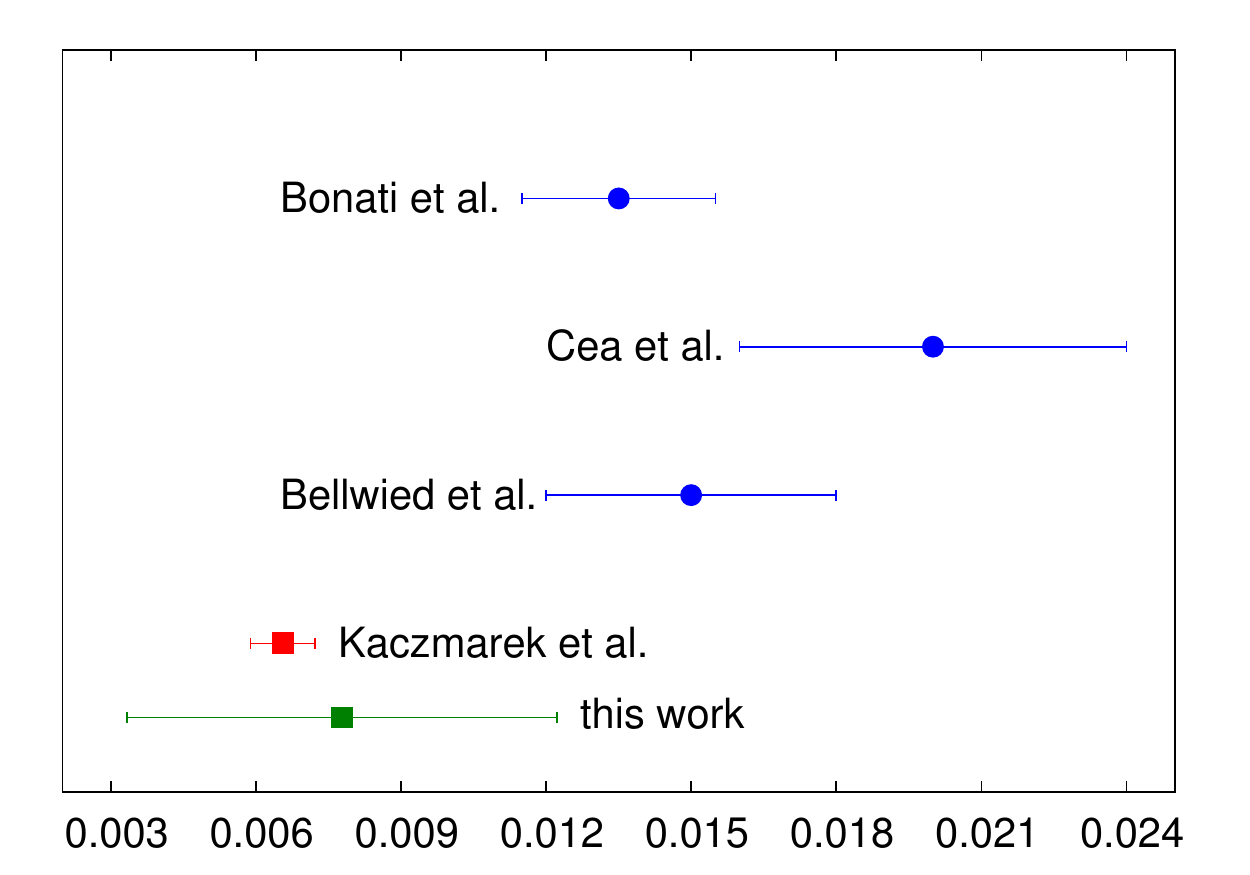}
\caption{\small (Left) $2\kappa_{ll}f'_G(z)$, obtained from rearranging Eq.~\protect\eqref{eq:mixed}. (Right) This result compared to other lattice results for $\kappa_{ll}$ (These are, from bottom to top, \cite{Kaczmarek:2011zz}, \cite{Endrodi:2011gv}, \cite{Cea:2014xva} and  \cite{Bonati:2014rfa}). The other results are for $\kappa_B$, the curvature w.r.t. the baryochemical potential $\mu_B$, therefore we have divided our result for $\kappa_{ll}$ by a factor of 9 (see text).}
\label{fig:kll}
\end{figure}

Fig.~\ref{fig:kll} shows our current results for the curvature $\kappa_{ll}$. Our eventual goal is  to calculate the mixed susceptibilities, Eq.~\eqref{eq:mixed} to a good precision for all quark masses over a wide range of $z$. This will allow us to map out the shape of the scaling function $f_G'(z)$. The curvatures can then be formally obtained from a one-parameter fit to Eq.~\eqref{eq:mixed}. Unfortunately, with our current range of temperatures we are restricted to the $z$-interval $-1\lesssim z \lesssim 1$. We therefore merely varied $\kappa_{ll}$ by hand to get upper and lower bounds on $\kappa_{ll}$. We show our results on the left-hand side of Fig.~\ref{fig:kll} (Left); Fig.~\ref{fig:kll} (Right) compares our result to the values obtained from other calculations.

\begin{figure}[!ht]
\includegraphics[width=0.45\textwidth]{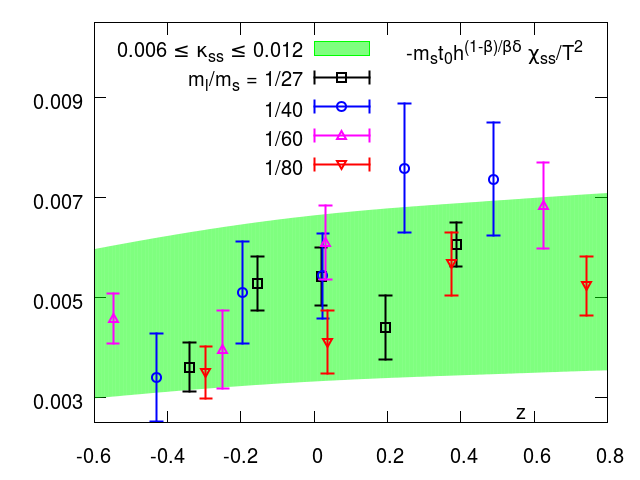}%
\hspace{0.10\textwidth}%
\includegraphics[width=0.45\textwidth]{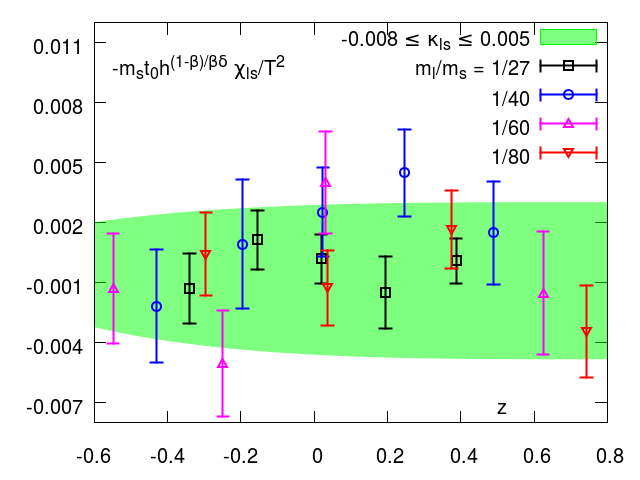}
\caption{\small Similar to Fig.~\protect\ref{fig:kll}, but now $2f_G'(z)$ times the strange and light-strange curvatures $\kappa_{ss}$ (left) and $\kappa_{ls}$ (right).\label{fig:kss}}
\end{figure}

Finally, Fig.~\ref{fig:kss} shows our results for the other two coefficients, namely $\kappa_{ss}$ and $\kappa_{ls}$. We see that both these coefficients are about an order of magnitude smaller than $\kappa_{ll}$. This would mean that the curvature along the baryon chemical potential $\mu_B$ direction, $\kappa_B \simeq \kappa_{ll}/9$ to a very good approximation. We also note that, within our current errors, the off-diagonal curvature $\kappa_{ls}$ could have either sign and could also possibly be zero although a slightly negative value seems to be preferred by the data.

\section{Conclusions}
\label{sec:conclusions}
The curvature of the QCD chiral phase transition line with $\mu$ is of interest both to theory as well as to the search for the critical point of QCD. At this conference, we presented preliminary results for the light quark curvature, calculated using HISQ fermions on $N_\tau=6$ lattices. Furthermore, by including the strange chemical potential as well, we were able to calculate the other elements of the $2\times2$ curvature matrix $\mathbf{K}$. To the best of our knowledge, these other coefficients have not been studied before. The light curvature is the predominant element of this matrix, the other elements being about an order of magnitude smaller. Currently, the least constrained matrix element of our calculation is the off-diagonal element $\kappa_{ls}$ which could have either sign, although the data seem to prefer a slightly negative value.

\section*{Acknowledgements}
The ensembles used in this calculation were produced at Jefferson Laboratory in the United States and the measurements were carried out on the Tianhe-I and II machines in China. PH is partly supported by Research Fund No. 11450110399 for International Young Scientists from the National Natural Science Foundation of China.

\end{document}